# An Informatics Software Stack for Point Defect-Derived Opto-Electronic Properties: The Asphalt Project

Jonathon N. Baker
Department of Materials Science and Engineering, 911 Partners Way, Suite 3002, North Carolina State University, Raleigh, NC 27695, USA

Preston C. Bowes
Department of Materials Science and Engineering, 911 Partners Way, Suite 3002, North Carolina State University, Raleigh, NC 27695, USA

Joshua S. Harris
Department of Materials Science and Engineering, 911 Partners Way, Suite 3002, North Carolina State University, Raleigh, NC 27695, USA

Douglas L. Irving*
Department of Materials Science and Engineering, 911 Partners Way, Suite 3002, North Carolina State University, Raleigh, NC 27695, USA
* Corresponding author. E-mail: dlirving@ncsu.edu

**Abstract:**
Computational acceleration of performance-metric-based materials discovery via high-throughput screening and machine learning methods is becoming widespread. Nevertheless, development and optimization of the opto-electronic properties that depend on dilute concentrations of point defects in new materials have not significantly benefited from these advances. Here, we present an informatics and simulation suite to computationally accelerate these processes. This will enable faster and more fundamental materials research, and reduce the cost and time associated with the materials development cycle. Analogous to the new avenues enabled by current first-principles-based property databases, this type of framework will open entire new research frontiers as it proliferates.

**Suggestion for Graphical abstract**: Fig. 2.

I. INTRODUCTION

In 2011, the National Science and Technology Council (NSTC) announced the Materials Genome Initiative for Global Competitiveness. [1] This white paper outlined the 7 stages of materials development, which has traditionally taken 10 to 20 years from initial investigation of a material to its first industrial use. These stages are shown in Fig. 1. The NSTC envisioned a massive collaboration to accelerate all stages, which would enable data and algorithms to be shared among computational and experimental researchers and industry.

Thus far, the vast majority of effort has yielded tools aimed at computationally accelerating the first three stages with a general focus on bulk property optimization. These efforts resulted in the creation of repositories like the Materials Project [2], OQMD [3], and AFLOW [4], which have built up vast databases of bulk material properties. These repositories can be screened to find good candidate materials for a given application, or data mined and used with artificial intelligence algorithms to predict new materials with desired properties. [5,6] This has resulted in a number of significant advances in areas such as rechargeable batteries, thermoelectrics, and photovoltaics. [7,8]

Nevertheless, much of the original vision and promise of the materials genome initiative has yet to materialize in systems where the properties of interest depend on dilute concentrations of point defects, as in opto-electronics. This is even more problematic considering that stages 2 and 3 are the most time consuming and expensive steps for optimizing such properties, and are most critical for taking a candidate material system from the laboratory to commercial products and devices.

This is partially an architectural problem, and partially a cost-benefits problem, with the two facets exacerbating one another. For many applications it is sufficient to focus the optimization on bulk properties. From a computational perspective, this is desirable as these properties can be captured with a prescribed set of simulations per material and usually utilize smaller and less computationally expensive simulations as compared to those required for point defects. Such efforts then focus on assessing the properties of thousands of materials in a satisfactory time frame and at a high enough level of theory. Because of the nature of these calculations, the data frameworks developed to handle them evolved to handle a fixed set of simulations for each material that are sufficient to predict target properties.

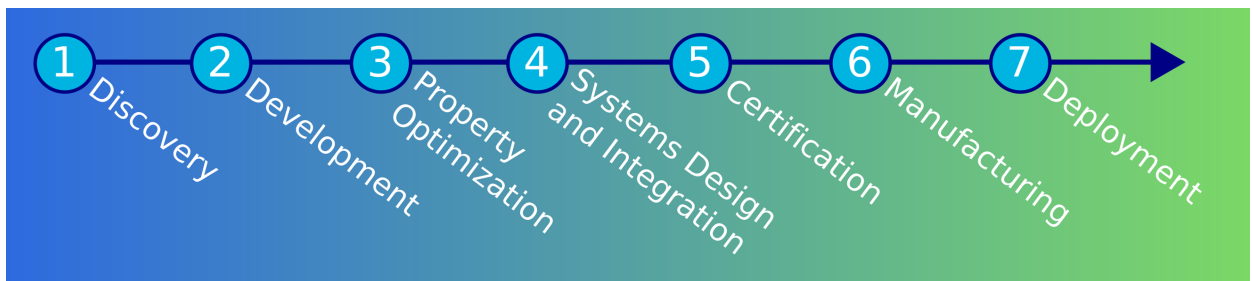

*Figure 1 The Materials Development Pipeline - Adapted from Holdren* et al.*[1]*

Prediction of point defect dependent properties needed for development of optical and electronic devices does not map well into this framework and is computationally expensive. Electrical properties dependent on doping can vary over several decades and are controlled by the intentional and unintentional point defects present. Larger bandgaps than that of Si are often needed for optical and high power electronics. Doping in larger bandgap systems is more complicated as defects exhibit deeper trap or higher charge states and defects may prefer to associate or complex. As a result, compensation of intentional impurities by a host of unintentional defects is common, leading to deviations from doping rules honed in the development of Si technology. Further complicating matters is that this information cannot be obtained from a fixed, prescribed set of calculations like the bulk properties.

The most common technique for examining and reporting point defects is the formation energy diagram, which plots the defect formation energies at specific chemical potentials as a function of Fermi level. While this provides information on the likelihood of formation and the ionization states of the point defects, they do not map naturally into experimentally observable properties, making them inefficient for design. This is due to the fact that the ensemble of defects present collectively controls the position of the Fermi level and, as a result, the constituent defects are coupled. Collectively, this phenomenon can be studied through a grand canonical thermodynamics ensemble, but this leads to a massive combinatorial space of point-defect-derived properties for each material. Because of these factors, it is a challenge to translate collections of point defect simulations to material properties, especially with frameworks designed around prescribed sets of calculations. This is the aforementioned architectural problem.

In addition to the above, the underlying point defects must also be simulated at a more computationally expensive level of theory than most of the bulk properties, as traditional functionals are known to under predict the materials bandgap and, thus, the defect thermodynamic transition levels. At the same time, the simulations themselves must be in larger supercells than those used for obtaining bulk properties to isolate the defects from their periodic images, and many charge states of each defect need to be simulated for each material. This leads to the cost-benefits problem: simulate thousands of materials with a focus on bulk properties, or only simulate one or two materials at a deep enough level to predict their functional electrical properties?

These two factors have led to the current evolution of high throughput methods: there are several mature frameworks and data repositories for computationally accelerating discovering and optimizing bulk properties, and far fewer for computationally accelerating materials development and property optimization related to point defects. This article discusses development of a point defects informatics framework intended to address this pressing need. The framework is designed around high throughput simulation and postprocessing of point defects, backed by a persistent database, to provide input to a grand canonical thermodynamics model. Ultimately, this allows direct investigation of the relationship between processing, point defects, and the macroscopic electrical properties and optical responses.

The remainder of the article discusses some practical considerations associated with these models, such as the scale of computational resources required, then discusses the different steps in the framework's workflow. While this framework was under development, others in the community began concurrently developing their own, such as pyCDT [9], although the different focus areas of the projects may ultimately lead to complementary roles in the community. This will be briefly discussed. Finally, the article closes with highlights from recent investigations made possible by this framework and future steps enabled by it.

## II. PRACTICAL CONSIDERATIONS AND RESOURCE REQUIREMENTS

The methodology for translating point defect simulations to material properties can be quite involved, and is discussed in depth elsewhere. [10-14] The most salient point from the perspective of informatics systems design is this: the more energetically favorable defects naturally have high concentrations, while the unfavorable defects naturally have low concentrations. This encourages exploration of many possible defect configurations, since leaving out an energetically favorable defect can affect the accuracy of the thermodynamics simulations and property prediction, while including even very many unfavorable defects has no real impact on macroscopic properties.

The simulations used for this type of work utilize density functional theory (DFT) calculations with hybrid exchange-correlation functionals. Specifically, DFT calculations using the Heyd, Scuseria, and Ernzerhof (HSE) functional are currently preferred for point defect and electronic structure simulations versus those using functionals based purely on the generalized gradient approximation (GGA), such as the widely used Perdew-Burke-Ernzehof (PBE) functionals. [15,16] While calculations based on GGA and PBE functionals accurately reproduce many of the properties relevant to performance-based material selection, they have shortcomings when used to examine the bandgap and charge localization, and by extension, defect in-gap state levels, leading to uncertainty in calculated point defect properties. [17] These shortcomings are believed to be due mainly to electron self-interaction. [17]

The drawback of the HSE functional is its high computational cost, which is typically more than an order of magnitude higher than calculations using traditional functionals. [18] Compounding this is the need for thorough exploration of potential isolated and complexed (associated) point defect configurations. This makes obtaining a dataset large enough to make meaningful predictions about material properties quite expensive.

The number of required simulations for a given material depends sensitively on its technical maturity and defect complexity. Electronic grade materials often require fewer impurities to be simulated for complete coverage because they are often of higher purity. Even with the high purity and lower number of impurities, there can be a host of configurations and defect complexes that ultimately govern macroscopic properties. With that said, many important electronic components do not have the same level of purity requirements because source purification can significantly increase the cost of each component. While some impurities are benign, this often cannot be determined *a priori*.

As an example of the challenges in each class of material, a researcher studying electronic grade donor-doped AlN may need to only consider the donor (silicon), native defects (vacancies), and common unintentional impurities (carbon and oxygen). Silicon in AlN, however, is stable in over 100 symmetrically inequivalent configurations, each in multiple charge states, due to its propensity to form multimember defect complexes. [13] In contrast, an electroceramic system like bulk grown $SrTiO_3$ typically has dozens of impurities present, but they may only be favorable in three or four isolated and complexed configurations, requiring only tens of simulations to fully characterize each impurity, but with a large number of impurities needed to determine certain properties. [11]

This base layer of point defect simulations must be supplemented with simulations of thermodynamically relevant phases (prime and competing) in order to construct the chemical potential space and the electronic structure to obtain various electronic properties such as the bandgap, valance band maximum, effective masses, and the density of states. Additional calculations are also typically performed to obtain predicted defect spectroscopic signatures. This can amount to thousands of simulations and hundreds of thousands to millions of datapoints for each material. The need to intelligently and quickly interact with this scale and depth of data motivated development of the high throughput application stack, illustrated in Fig. 2, which enables efficient setup of and interaction with point defect DFT simulations.

## III. ASPHALT PROJECT APPLICATION STACK

The application suite consists of the informatics application itself (asphalt), its accompanying database (bitumen), a tool for settings up new point defects simulations quickly, and a set of simulation tools based on grand canonical point defect thermodynamics (defect concentration solver). In keeping with the Unix design philosophy, these components are only weakly coupled, and, where necessary, communication between the applications is via plain text document. This is intended to make it simpler to deploy and maintain the software, to retool it if it becomes necessary, or to more simply swap in another tool for one of the components if desired.

These programs are currently tooled to support a specific DFT application, the Vienna Ab Initio Simulation Program (VASP) [19], due to its large install base, although the toolset's design is flexible and retooling to support additional programs is straightforward.

Additionally, user-facing functionality is presented through executables and direct database interaction, rather than through a web interface or APIs (application programming interface). This dramatically lowers the learning curve for the asphalt informatics stack, since the end-user accessible features have easy-to-access, succinctly documented ways of being used without needing to write a program or script. More complex functionality is of course still accessible via the API, which carries the same time-investment requirements here as for any other project with an API, but most of the use cases will be covered by existing functionality in the executables or simple queries of the database.

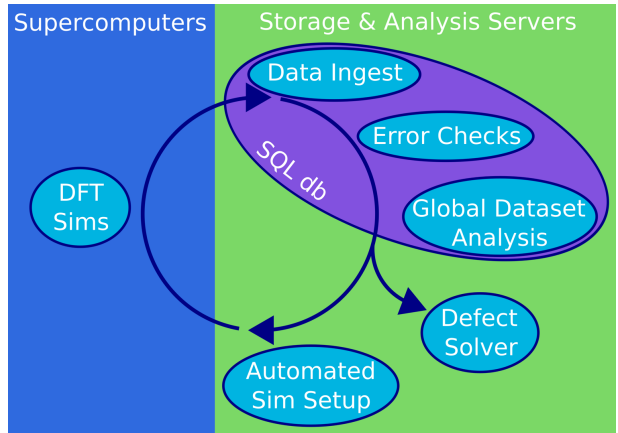
*Figure 2 High throughput point defects workflow.*

A. Asphalt and Bitumen

As with most large database systems, this informatics system is composed of an application to process and ingest the data (asphalt) paired with a persistent database to store and make the data available (bitumen). Bitumen is implemented in SQL due to the relational structure of the dataset and the format's industry-wide usage and support, and currently runs on sqlite. Due to its mature ecosystem, low cognitive impedance, and the ease of concisely expressing complex algorithms, asphalt is written in Common Lisp.

Together, they provide a bespoke solution to the first part of the architectural problem discussed in the introduction: they provide a method for automatically postprocessing and persistently storing arbitrary numbers and types of point defect simulations, and associated calculations. Crucially, the ingest and classification process is designed to operate with minimal researcher interaction and almost no prior knowledge of the data it is ingesting.

Because of this ability to automatically sort and classify the simulations into different materials and defect types based only on the simulations themselves, the process for importing new data and legacy data generated without this framework are the same. This means that this framework can easily ingest and postprocess existing datasets, rather than needing to generate new ones or to manually create a metadata file detailing the simulation's types and relationships. This also makes it compatible with datasets generated with other tools.

Importing data with asphalt is simple. When the program is invoked with the *import* option and a directory containing simulations, it records the file/directory structure and parses data from all unique files. This data is converted to a simulation level representation when asphalt is invoked with the *analyze* option. Lastly, after a short step to tag the electronic structure simulation and any optical transition calculations, the defect analysis step identifies the types of and relationships between simulations in the dataset and runs one of the modularly integrated finite-size correction codes, which correct the formation energy to the dilute limit for smaller supercells. All of this information is made available in dashboard-style tables in the database. The data can then be exported into more portable csv-based formats for import into defect concentration solver based tools or chemical potential space mapping utilities, or the data can be directly used by other programs via the SQL interface.

A combination of error checking routines, the automatic import and classification process, and the way the data is presented means that many common errors are either automatically detected or appear as obvious pattern deviations. This is important, as tedious and repetitive tasks performed or set up by humans tend to accumulate errors, and this is magnified when multiple researchers are collecting the dataset over many months or years. With the plethora of data in even a single simulation within such a dataset, these errors easily go unnoticed, affecting the accuracy of any models and conclusions based on the data.

Because of its architecture as a standalone executable and database file, deployment of the database is flexible. It can be deployed as a single monolithic database for handling multiple material systems, or as smaller instances for each material. Pragmatically, deployment as smaller instances makes the data more portable because the database files are smaller without losing the ability to agglomerate the data later in a data warehouse, and also makes it easier to focus on the data for a single material system when checking for errors or inconsistencies.

B. Defect Concentration Solver

The defect concentration solver is a point defects simulation tool based on the grand canonical thermodynamics ensemble. It addresses the second part of the architectural problem discussed in the introduction, by providing a way to explore macroscopic opto-electronic properties derived from point defect concentrations in the vast combinatorial space of processing and doping. Techniques for calculating equilibrium point defect concentrations in semiconductors and insulators with known or assumed impurity chemical potentials are not new, and implementations are becoming more widely available.[9,20]

This implementation provides several key improvements. First, the chemical potentials of dopant species are not a natural representation for the amount of dopant, but the existing implementations require it as a coordinate. The exact value of both the impurity and native chemical potentials affects the distribution amongst the various isolated and complexed forms. This limits the practical utility of parametric studies or studies with the impurity chemical potential determined by the solubility limiting phase, as are often done. This implementation accepts either impurity chemical potential or impurity concentrations.

Second, real materials are rarely measured or used in devices at temperatures high enough for full point defect equilibrium. The measurement/operational temperature is usually some lower temperature with the point defects in a state of constrained equilibrium. This implementation calculates both the high temperature, fully equilibrated populations of electrons, holes, and point defects, as well as the low temperature populations in constrained equilibrium. Using the point defects data from the database and chemical potentials of the native materials set by the processing environment, these populations are calculated self consistently by solving for the chemical potentials and Fermi level at the annealing or growth temperature, and then allowing ionic charge redistribution to maintain charge neutrality at the quench temperature. This core simulation library is leveraged to create a virtual electronic materials workbench, where many opto-electronic properties can be explored *in silico*.

In principle, any optical or electronic property that arises from point defect or charge carrier populations can be explored with programs building on this core functionality. Programs already exist for some of the most common envisioned use cases. The variation of point defect and charge carrier concentrations with processing environment can be explored for a fixed impurity profile. All permutations of a set of impurity concentration permutations can be simulated in a single-shot run. Additionally, single-anneal multi-quench simulations can be used to explore thermal activation processes. Data from these tools, and variations on them, can be loaded into further postprocessing tools to explore properties such as conductivity, gas solubility as a function of impurity and processing, and modification of observable electrical properties due to contact diode effects. Combining the calculated concentrations with optical signatures from the database or other sources allows exploration of optical absorption/emission signatures with processing and doping.

C. Semi-automated Simulation Setup Tool

The toolset for setting up new simulations is a set of python scripts which call a library designed around DFT software input file manipulation. Although this process has been heavily automated, it is intentionally not fully automated, and is not directly coupled with the database. Rather, the digested simulation data is post-processed with the defect concentration solver to examine defect favorability and identify missing charge states, and then that information is used by the researcher to set up the next round of simulations.

The reason for not fully automating this process is that, while every material system has the same general types of point defects, many materials have defect configurations which are uniquely favorable in that system, and unfavorable in most others. As an example, consider the hundreds of multi-member $v_{Al}$-$nSi_{Al}$ complexes in AlN. [13] It would be both wasteful and prohibitively expensive to simulate all such defects for all materials at the required level of theory. Leaving control of this process with the researcher allows them to leverage their expertise with their particular material system to realize a substantial reduction in computational expense.

The python scripts for this tool accept simple input files to specify common tasks such as modifying primitive cells into supercells, adding defects and defect complexes, setting atomic relaxation flags to specify relaxation spheres around defects, and setting up defect calculations across a set of charge states. The scripts also make it simple to set up additional simulations which are consistent with the numeric settings of an existing dataset, via functionality for loading and modifying existing simulations. This level of automation allows dozens of simulations to be set up in a matter of minutes while still providing full control to the researcher.

This component of the software stack is most similar in scope to the pyCDT project. [9] As noted in the introduction, these tools were developed independently and concurrently due to a common need to automate defect simulation set up. However, pyCDT did not yet support creation of defect complexes or optical transition simulations when its accompanying article was published, and these areas may still be under development. Due to the loose coupling between framework components, substitution of this tool subset with pyCDT or any other high throughput simulation setup tool is straightforward if needed or desired.

## IV. WHAT DOES THIS ENABLE?

This software stack has been developed and refined for many years, and over time its capabilities have grown. As these capabilities have grown, they have been applied to and helped resolve real world materials science challenges. This section discusses highlights of some of this research.

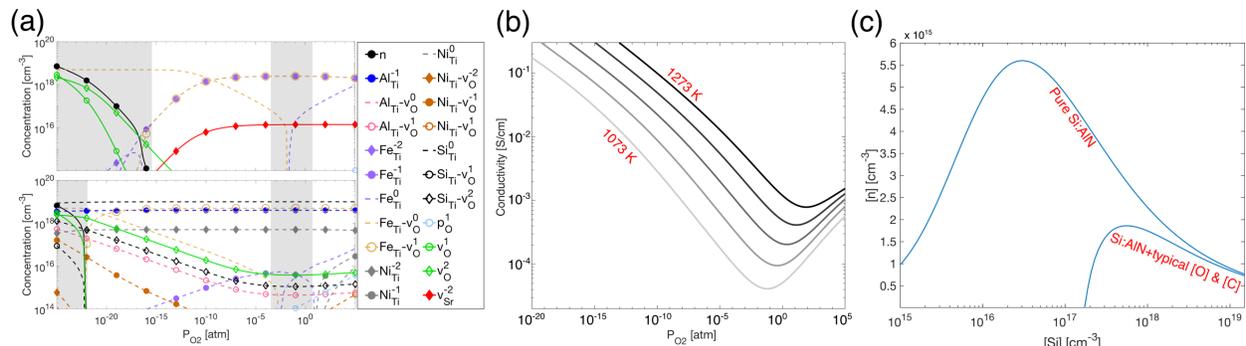

*Figure 3 Highlights from work enabled by this framework. (a) Coloration of Fe:SrTiO$_3$. (b) Temperature dependence of SrTiO$_3$ conductivity with an impurity ensemble. (c) Si compensation knee in AlN*

### A. Fe-doped SrTiO3

Strontium titante (SrTiO$_3$) is widely regarded as a prototypical oxide perovskite. Iron is a commonly studied impurity in this system, as it reversibly switches between clear and brown in response to changing system oxidation, providing a simple probe of the electrochemical state of the system. In the past, combined EPR and optical absorption measurements provided indirect evidence that the neutral $Fe_{Ti}$ (iron substituting for titanium) defect was associated with the brown coloration, which manifests as broad spectroscopic absorption shoulders centered near 460 nm (2.7 eV) and 570 nm (2.18 eV). [21] Defect reaction models developed to describe iron in SrTiO3, largely based on high temperature conductivity data, invoked $Fe_{Ti}^0$, $Fe_{Ti}^{-1}$, n, p, and the +2 oxygen vacancy ($v_O^{+2}$), and ignored inter-defect interactions. [22]

However, additional EPR measurements from the same body of work detected non-negligible concentrations of strongly interacting first-nearest-neighbor iron oxygen vacancy complexes. [21] This informatics framework and defect concentration solver were used to resolve the dichotomy and establish a comprehensive model that included the effects of the $Fe_{Ti} - v_O$ complex. [10]

Counting both calculations for thermodynamic data and optical data, more than a hundred simulations were performed spanning the full range of possible native defects, iron, and its complexes. Additional simulations spanning sulfur, chlorine, silicon, aluminum, sodium, magnesium, barium, and nickel impurities were also available. This dataset was pared down based on absorption energy, defect favorability, and spin selection.

The only two remaining absorptions after paring down were associated with excitations from $\text{Fe}_{\text{Ti}}^0$, at 2.72 and 2.17 eV, a nearly perfect match for the experimental absorption. And, with the inclusion of the complex in the thermodynamics model, there was a rapid onset and gradual increase of $\text{Fe}_{\text{Ti}}^0$ at a $P_{O2}$ consistent with experiment, as shown in Fig. 3(a).

The additional impurity simulations were used together with simulated variations in processing temperature and processing chemical potentials to explore the robustness of the conclusions. The coloration onset location had minor sensitivity to temperature and processing environment, and it was found that the coloration onset was typically maintained until the iron concentration was exceed by a factor of between 2 and 8, depending on the codopant.

B. High temperature conductivity

The carrier concentrations and conductivity, however, were found to be extremely sensitive to even background impurity content. This was at odds with the models typically invoked for explaining conductivity in this type of material, which ascribe all of the behavior to a single dominant impurity, and neglect background impurities. This was explored in work by Bowes et al..[11]

This work compared predictions against a classic experimental paper in the oxide perovskite literature, Chan and Smyth's *Nonstoichiometry in SrTiO$_3$*[23] This article was selected as it is one of the few in the field to report an impurity profile with its conductivity results.

Specifically, 9 of the 17 reported impurities, spanning hundreds of simulations, were sequentially added in order of decreasing concentration to a conductivity model to explore the emergent properties of the impurity ensemble. It was found that the ensemble behaved completely different than the individual impurities, and that as more impurities were added the behavior more closely approximated the reported experimental data and trends in temperature as shown in Fig. 3(b). These results indicate a need for caution in applying the traditional single-impurity defect reaction models in these materials, especially without impurity concentrations, as secondary dopants can substantially modify the behavior.

C. Donor compensation in AlN

The informatics framework discussed herein has also enabled substantial mechanistic insights in III-V systems. In AlN and GaN, n-type doping is typically accomplished via silicon doping. However, while silicon in GaN exhibits nearly ideal donor doping all the way into the electronically degenerate doping limit, silicon in AlN and Al-rich AlGaN compositions suffers from a "compensation knee" before higher electron concentrations are realized.

Essentially, as the silicon content is increased, there is initially a steady increase of electrons with silicon, but this relationship breaks down and even reverses slope at higher silicon content. Several extrinsic and intrinsic mechanisms have been proposed over time to explain this behavior, including the silicon DX state [24], aluminum and gallium vacancies [25], silicon substitution on the nitrogen site [25], and silicon complexes [26,27].

Using this informatics stack, we performed an extensive investigation of this phenomenon, and found that the compensation knee actually arises from a form of self-compensation, via the complexes originally hypothesized by Uedono *et al.* [26]. As the silicon content increases, it initially all incorporates as isolated $\text{Si}_{\text{Al}}$ and $\text{Si}_{\text{Al}}$-dx, leading to the steady increase in electron concentration.

However, as the favorability of the single silicon defect increases, the favorability of multimember silicon complexes increase at multiplicatively higher rates. Eventually, the one, two, and three member complexes begin to affect and then control the bulk charge compensation, leading to the knee shape. Due to the large number of geometrically inequivalent configurations, this required performing several hundred simulations just to capture silicon. Hundreds more simulations were performed for carbon and oxygen and their respective complexes, which tend to clamp the knee onset, effectively increasing the silicon concentration of the peak, as shown in Fig. 3(c). Matches between experimental measurements on different faces of the knee and predicted changes in the optical properties of other impurities in the system as they change their dominant forms with the changing compensation lend additional support to the mechanism. [13]

## V. CONCLUDING REMARKS AND ACKNOWLEDGEMENTS

With the convergence of large amounts of computational power, sufficiently accurate DFT calculations to inform more sophisticated thermodynamics models, and the advent of high throughput informatics frameworks to facilitate and integrate all of the calculations, the time to take advantage of these techniques is now. It is envisioned that this informatics framework, and other frameworks like it, will allow more researchers to computationally explore the interplay of point defects and functional properties in their material systems while leveraging their existing data. This will substantially accelerate development of new and existing material systems from the laboratory to commercial applications and could be used to address a variety of scientific and industrial problems in multiple sectors.

This also enables some interesting next steps. Machine learning requires large amounts of structured data to operate on for accuracy. In large part, the advances in applying machine learning and artificial intelligence (AI) to materials have been enabled via first-principles-based bulk property repositories like AFLOW, OQMD, and the Materials Project. As the technology for building up point defect property datasets proliferates and becomes more widespread, it will begin opening up exploration and data mining of high accuracy point defect energies, complementing the existing AI work on more bulk-focused properties. It also opens up an additional frontier which has been neglected until now: direct application of AI to the functional electrical properties of materials as a function of their doping and processing. This is in many ways a more economically relevant issue than AI application to the point defect energies themselves, and would further reduce the cost and time associated with the fundamental materials research involved in the development and property optimization stages of opto-electronic materials.

Finally, structured and reliable data on point defect behavior will enable the next generation of device simulation. One of the biggest challenges of current generation device simulation is a lack of accurate data on point defect behavior. Those products which can deal with arbitrary point defects often require input of both the defect state levels and the defect concentration, which can be difficult or impossible to acquire experimentally for situations where a property is derived from an ensemble of related defects instead of a single one. This informatics stack, and others like it, can directly inform such applications, providing an integrated set of data and applications spanning and accelerating stages 2 through 4 of the materials development pipeline.

The authors gratefully acknowledge support from the AFOSR via grants FA9550-14-1-0264 and FA9550-17-1-0318, which are in the Aerospace Materials for Extreme Environments program of Dr. Ali Sayir. PCB acknowledges support from the NDSEG. Computer time was provided for the generation of data by NERSC and DoD HPCMP.

[1] J. P. Holdren et al., National Science and technology council OSTP. Washington, USA (2011).
[2] A. Jain, S. P. Ong, G. Hautier, W. Chen, W. D. Richards, S. Dacek, S. Cholia, D. Gunter, D. Skinner, G. Ceder, and K. A. Persson, APL Materials **1**, 011002 (2013).
[3] J. E. Saal, S. Kirklin, M. Aykol, B. Meredig, and C. Wolverton, Jom **65**, 1501 (2013).
[4] S. Curtarolo, W. Setyawan, G. L. Hart, M. Jahnatek, R. V. Chepulskii, R. H. Taylor, S. Wang, J. Xue, K. Yang, O. Levy, et al., Computational Materials Science **58**, 218 (2012).
[5] W. Ye, C. Chen, S. Dwaraknath, A. Jain, S. P. Ong, and K. A. Persson, MRS Bulletin **43**, 664 (2018).
[6] C. Toher, C. Oses, J. J. Plata, D. Hicks, F. Rose, O. Levy, M. de Jong, M. Asta, M. Fornari, M. B. Nardelli, et al., Physical Review Materials **1**, 015401 (2017).
[7] A. Jain, Y. Shin, and K. A. Persson, Nature Reviews Materials **1**, 15004 (2016).
[8] K. Alberi, M. B. Nardelli, A. Zakutayev, L. Mitas, S. Curtarolo, A. Jain, M. Fornari, N. Marzari, I. Takeuchi, M. L. Green, M. Kanatzidis, M. F. Toney, S. Butenko, B. Meredig, S. Lany, U. Kattner, A. Davydov, E. S. Toberer, V. Stevanovic, A. Walsh, N.-G. Park, A. Aspuru-Guzik, D. P. Tabor, J. Nelson, J. Murphy, A. Setlur, J. Gregoire, H. Li, R. Xiao, A. Ludwig, L. W. Martin, A. M. Rappe, S.-H. Wei, and J. Perkins, Journal of Physics D: Applied Physics **52**, 013001 (2019).
[9] D. Broberg, B. Medasani, N. E. Zimmermann, G. Yu, A. Canning, M. Haranczyk, M. Asta, and G. Hautier, Computer Physics Communications **226**, 165 (2018).
[10] J. N. Baker, P. C. Bowes, D. M. Long, A. Moballegh, J. S. Harris, E. C. Dickey, and D. L. Irving, Applied Physics Letters **110**, 122903 (2017).
[11] P. C. Bowes, J. N. Baker, J. S. Harris, B. D. Behrhorst, and D. L. Irving, Applied Physics Letters **112**, 022902 (2018).
[12] J. N. Baker, P. C. Bowes, J. S. Harris, and D. L. Irving, Journal of Applied Physics **124**, 114101 (2018).

[13] J. S. Harris, J. N. Baker, B. E. Gaddy, I. Bryan, Z. Bryan, K. J. Mirrieless, R. Collazo, Z. Sitar, and D. L. Irving, Applied Physics Letters **112**, 152101 (2018).
[14] J. N. Baker, P. C. Bowes, and D. L. Irving, Applied Physics Letters **113**, 132904 (2018).
[15] J. Heyd, G. E. Scuseria, and M. Ernzerhof, The Journal of Chemical Physics **118**, 8207 (2003).
[16] J. P. Perdew, K. Burke, and M. Ernzerhof, Physical review letters **77**, 3865 (1996).
[17] C. Freysoldt, B. Grabowski, T. Hickel, J. Neugebauer, G. Kresse, A. Janotti, and C. G. Van De Walle, Reviews of Modern Physics **86**, 253 (2014).
[18] V. L. Chevrier, S. P. Ong, R. Armiento, M. K. Chan, and G. Ceder, Physical Review B **82**, 075122 (2010).
[19] G. Kresse and J. Hafner, Physical Review B: Condensed Matter and Materials Physics **47**, 558 (1993).
[20] C. G. Van de Walle, D. Laks, G. Neumark, and S. Pantelides, Physical Review B **47**, 9425 (1993).
[21] K. Mueller, T. Von Waldkirch, W. Berlinger, and B. Faughnan, Solid State Communications **9**, 1097 (1971).
[22] T. Baiatu, R. Waser, and K.-H. Haerdtl, Journal of the American Ceramic Society **73**, 1663 (1990).
[23] N.-H. Chan, R. Sharma, and D. M. Smyth, Journal of The Electrochemical Society **128**, 1762 (1981).
[24] F. Mehnke, T. Wernicke, H. Pingel, C. Kuhn, C. Reich, V. Kueller, A. Knauer, M. Lapeyrade, M. Weyers, and M. Kneissl, Applied Physics Letters **103**, 212109 (2013).
[25] Y. Taniyasu, M. Kasu, and N. Kobayashi, Applied Physics Letters **81**, 1255 (2002).
[26] A. Uedono, S. Ishibashi, S. Keller, C. Moe, P. Cantu, T. Katona, D. Kamber, Y. Wu, E. Letts, S. Newman, S. Nakamura, J. S. Speck, U. K. Mishra, S. P. DenBaars, T. Onuma, and S. F. Chichibu, Journal of Applied Physics **105**, 054501 (2009).
[27] I. Bryan, Z. Bryan, S. Washiyama, P. Reddy, B. E. Gaddy, B. Sarkar, M. H. Breckenridge, Q. Guo, M. B. Graziano, J. Tweedie, S. Mita, D. L. Irving, R. Collazo, and Z. Sitar, Applied Physics Letters **112**, 062102 (2018).